\documentclass[onecolumn,12pt,draftclsnofoot]{IEEEtran}

\usepackage{mathrsfs}
\usepackage{mathptmx}
\usepackage{amsmath}
\usepackage{amsfonts}
\usepackage{amssymb}
\usepackage{graphicx}
\usepackage{subfigure}

\newtheorem{thm}{Theorem}

\newtheorem{cor}{Corollary}
\newtheorem{rem}{Remark}
\newtheorem{lem}{Lemma}

\newtheorem{assum}{Assumption}
\def\Ds{\displaystyle}

\def\diag{\mathrm{diag}}
\def\tp{\mathrm{T}}
\def\sgn{\mathrm{sign}}

\begin{document}

\title{A State Transition Matrix-Based Approach to Separation of Cooperations and Antagonisms in Opinion Dynamics}

\author{Deyuan Meng, Ziyang Meng, and Yiguang Hong 
\thanks{This work was supported by the National Natural Science Foundation of China (NSFC: 61473010, 61327807, 61320106006), the Beijing Natural Science Foundation (4162036), and the Fundamental Research Funds for the Central Universities (YWF-16-BJ-Y-27).}
\thanks{D. Meng is with the Seventh Research Division, Beihang University (BUAA), Beijing 100191, P. R. China, and also with the School of Automation Science and Electrical Engineering, Beihang University (BUAA), Beijing 100191, P. R. China (e-mail: dymeng@buaa.edu.cn).}
\thanks{Z. Meng is with the Department of Precision Instrument, Tsinghua University, Beijing 100084, P. R. China (e-mail: ziyangmeng@mail.tsinghua.edu.cn).}
\thanks{Y. Hong is with the Key Laboratory of Systems and Control, Institute of Systems Science, Chinese Academy of Sciences, Beijing, 100190, P. R. China (e-mail: yghong@iss.ac.cn).}
}


\date{}
\maketitle

\begin{abstract}
This paper is concerned with the dynamics evolution of opinions in the presence of both cooperations and antagonisms. The class of Laplacian flows is addressed through signed digraphs subject to switching topologies. Further, a state transition matrix-based approach is developed for the analysis of opinion dynamics, regardless of any assumptions on connectivity, structural balance or digon sign-symmetry of signed digraphs. It is shown that based on the separation of cooperations and antagonisms, a relationship can be bridged between opinion dynamics under signed digraphs and under conventional digraphs. This helps to solve convergence problems for opinion dynamics. In particular, bipartite consensus (or stability) emerges if and only if the associated switching signed digraph is simultaneously structurally balanced (or unbalanced), which generalizes the use of structural balance theory in opinion dynamics to the case study of changing network topologies.
\end{abstract}

\section{Introduction}\label{sec1}

The analysis of opinion dynamics has attracted considerable research interests recently because it plays an important role as the case study in characterizing behaviors for social networks. In general, the interactions among individuals are conveniently represented by conventional graphs with nonnegative weighted adjacency matrices when the cooperative interactions are only involved. It yields the class of so-called {\it conventional networks}, for which agreement or consensus has generated many notable results (see, e.g., \cite{ofm:07}-\cite{me:10} and references therein). However, for opinion dynamics, there may simultaneously exist antagonistic interactions in reality, especially when concerning competitive relations, such as love/hate, like/dislike and believe/disbelieve. This requires signed graphs, which admit positive and negative adjacency edge weights, to describe the simultaneous existence of cooperation and antagonism. As a consequence, a new class of {\it signed networks} emerges, of which there have been reported limited analysis approaches and results until now.

In the literature, one of the most considered ideas to perform analysis of opinion dynamics subject to antagonisms is how to find effective ways to reasonably connect signed networks with conventional networks. This helps benefit from well-developed results of conventional networks to solve convergence analysis problems of opinion dynamics in signed networks. In \cite{a:13}, the gauge transformation-based approach has been shown with the ability to solve bipartite consensus (or polarization) problems on signed networks by transforming them into the counterpart consensus problems on conventional networks. Such approach has been extensively applied in addressing bipartite consensus of general linear dynamics \cite{vm:14,zc:17}, interval bipartite consensus \cite{hz:14}-\cite{mdj:16} and bipartite containment tracking \cite{m:17}. In \cite{h:14}, a lifting approach has been introduced such that the opinion dynamics in signed networks can be remodeled as a corresponding class of opinion dynamics in conventional networks over a space of twice enlarged dimensions. It has been successfully applied in \cite{xcj:16} to study polarization of opinion dynamics in both cases of fixed and switching topologies. There have also been presented many other promising models for opinion dynamics in signed networks, see, e.g., \cite{msjch:16,pmc:16} for modulus consensus model, \cite{spjbj:15} for state-flipping model and \cite{al:15,jzc:17} for opinion forming model. Nonetheless, the existing models/approaches to opinion dynamics in signed networks generally require some restrictive assumptions, such as structural balance, digon sign-symmetry and strong connectivity.

In this paper, we are concerned with opinion dynamics when signed networks involve the class of Laplacian flows subject to switching topologies. By the separation of cooperative and antagonistic interactions, a state transition matrix-based approach is proposed to connect the opinion dynamics analysis in signed networks and in conventional networks, regardless of structural balance, connectivity and digon sign-symmetry. Consequently, it is technically and theoretically feasible to employ consensus results for conventional networks to deal with the convergence analysis of opinion dynamics in signed networks. Even though this advantage applies also to the lifting approach of \cite{h:14}, we resort to only the fundamental properties of the state transition matrices such that we can successfully overcome the drawback depending on the digon sign-symmetry to separate positive and negative interactions in opinion dynamics (for more details of this issue, see discussions made in \cite[Section VI, p. 2122]{h:14}). Furthermore, it is shown that bipartite consensus (respectively, stability) is achieved for opinion dynamics in a signed network if and only if the switching signed digraph associated with this network is simultaneously structurally balanced (respectively, unbalanced). We hence extend the application of the structural balance theory \cite{ch:56} to the general signed networks subject to switching topologies.

The remainder of our paper is organized as follows. We first end this section with basic notations. Then we give preliminary results of signed digraphs in Section \ref{sec2}. In Section \ref{sec3}, we take full advantage of the separation of cooperative and antagonistic interactions to present a state transition matrix-based approach to opinion dynamics in signed networks, based on which some convergence results are derived in Section \ref{sec4}. We finally make conclusions of our paper in Section \ref{sec7}.

{\it Notations:} Denote sets: $\mathbb{Z}_{+}=\{0,1,2,\cdots\}$, $\mathbb{Z}=\{1,2,3,\cdots\}$, $\mathcal{I}_{n}=\left\{1, 2, \cdots, n\right\}$, $\mathcal{Z}_{n}=\big\{A=\left[a_{ij}\right]\in\mathbb{R}^{n\times n}:a_{ij}\leq0,\forall i\neq j,\forall i,j=1,2,\cdots,n\big\}$ and $\mathcal{D}_{n}=\big\{D=\diag\{d_{1},d_{2},\cdots,d_{n}\}:d_{i}\in\{\pm1\},\forall i=1,2,\cdots,n\big\}$. Let us also denote $1_{n}=\left[1,1,\cdots,1\right]^{\tp}\in\mathbb{R}^{n}$, $\mathrm{int}[a]$ as the maximum integer not greater than scalar $a\in\mathbb{R}$ and $\Phi_{A(t)}(t,t_{0})$, $\forall t\geq t_{0}$ as the state transition (or fundamental) matrix of a linear system of the form $\dot{\xi}(t)=A(t)\xi(t)$. For any matrix $A=\left[a_{ij}\right]\in\mathbb{R}^{m\times n}$, we define four matrices $\left|A\right|=\left[|a_{ij}|\right]\in\mathbb{R}^{m\times n}$; $\Delta_{A}=\diag\left\{\sum_{j=1}^{n}a_{1j},\sum_{j=1}^{n}a_{2j},\cdots,\sum_{j=1}^{n}a_{mj}\right\}\in\mathbb{R}^{m\times m}$; $A^{+}=\left[a_{ij}^{+}\right]\in\mathbb{R}^{m\times n}$ with entries $a_{ij}^{+}=a_{ij}$ if $a_{ij}>0$ and $a_{ij}^{+}=0$ otherwise; and $A^{-}=-\left(-A\right)^{+}$. If $a_{ij}\geq0$ holds for all $i\in\mathcal{I}_{m}$ and $j\in\mathcal{I}_{n}$, then $A$ is called a nonnegative matrix, denoted by $A\geq0$. For a square nonnegative matrix $A\in\mathbb{R}^{n\times n}$, we call it a stochastic (or substochastic) matrix if $A1_{n}=1_{n}$ (or $A1_{n}\leq1_{n}$). If $A\in\mathbb{R}^{n\times n}$ is nonnegative, then we can define the Laplacian matrix associated with $A$ as $\mathcal{L}_{A}=\left[l_{ij}^{A}\right]\in\mathbb{R}^{n\times n}$, with its entries such that $l_{ij}^{A}=\sum_{k=1,k\neq i}^{n}a_{ik}$ if $j=i$ and $l_{ij}^{A}=-a_{ij}$ if $j\neq i$.

\section{Preliminaries of Signed Digraphs}\label{sec2}

\subsection{Fixed Topology}

A signed directed graph, or simply {\it signed digraph} for short, is denoted as a triple $\mathcal{G}=\left(\mathcal{V},\mathcal{E},\mathcal{A}\right)$, where $\mathcal{V}=\left\{v_{i}:i\in\mathcal{I}_{n}\right\}$, $\mathcal{E}\subseteq\mathcal{V}\times\mathcal{V}=\left\{(v_{i}, v_{j}):v_{i},v_{j}\in\mathcal{V}\right\}$ and $\mathcal{A}=\left[a_{ij}\right]\in\mathbb{R}^{n\times n}$ are node set, edge set and weighted adjacency matrix, respectively. In $\mathcal{G}$, $a_{ij}\neq0\Leftrightarrow(v_{j}, v_{i})\in\mathcal{E}$ and $a_{ij}=0$ otherwise for $i$, $j\in\mathcal{I}_{n}$, where $a_{ii}=0$ or $(v_{i},v_{i})\not\in\mathcal{E}$, $\forall i\in\mathcal{I}_{n}$ is considered. A directed edge from $v_{j}$ to $v_{i}$ is denoted as $(v_{j}, v_{i})\in\mathcal{E}$, under which $v_{j}$ is called a {\it neighbor} of $v_{i}$. Let $\mathcal{N}_{i}=\left\{j:(v_{j}, v_{i})\in\mathcal{E}\right\}$ represent the label set of all neighbors of $v_{i}$. If there exist distinct nodes $v_{i}$, $v_{k_{1}}$, $\cdots$, $v_{k_{l-1}}$, $v_{j}$ to form a finite sequence with $l$ ($1\leq l<n$) edges $\left\{(v_{k_{0}},v_{k_{1}}),(v_{k_{1}},v_{k_{2}}),\cdots,(v_{k_{l-1}},v_{k_{l}})\in\mathcal{E}:k_{0}=i, k_{l}=j\right\}$, then $\mathcal{G}$ is said to have a (directed) path from $v_{i}$ to $v_{j}$. We thus say that $\mathcal{G}$ is {\it strongly connected} if it has paths between each distinct pair of nodes.

Signed digraphs, in contrast to conventional digraphs, admit two additional properties in addition to connectivity: {\it structural balance} and {\it digon sign-symmetry} (\cite{a:13,ch:56,z:82}). We say that $\mathcal{G}$ is structurally balanced if it has a bipartition $\left\{\mathcal{V}^{(1)},\mathcal{V}^{(2)}\right\}$ of $\mathcal{V}$, where $\mathcal{V}^{(1)}\bigcup\mathcal{V}^{(2)}=\mathcal{V}$ and $\mathcal{V}^{(1)}\bigcap\mathcal{V}^{(2)}=\O$, such that $a_{ij}\geq0$, $\forall v_{i}$, $v_{j}\in\mathcal{V}^{(l)}$ for any $l\in\{1, 2\}$ and $a_{ij}\leq0$, $\forall v_{i}\in\mathcal{V}^{(l)}$, $\forall v_{j}\in\mathcal{V}^{(q)}$ for any $l\neq q$ and $l, q\in\{1, 2\}$; and $\mathcal{G}$ is structurally unbalanced, otherwise. If $a_{ij}a_{ji}\geq0$, $\forall i, j\in\mathcal{I}_{n}$, then $\mathcal{G}$ is digon sign-symmetric; and otherwise, $\mathcal{G}$ is digon sign-unsymmetric. In the literature (e.g., \cite{a:13,ch:56}), it shows that $\mathcal{G}$ is structurally balanced if and only if $D\mathcal{A}D=\left|\mathcal{A}\right|$ holds for some $D\in\mathcal{D}_{n}$; and $\mathcal{G}$ is structurally unbalanced, otherwise.

\subsection{Switching Topologies}

A time-varying signed digraph on the node set $\mathcal{V}$ is denoted by $\mathcal{G}(t)=\left(\mathcal{V},\mathcal{E}(t),\mathcal{A}(t)\right)$, in which the quantities corresponding to those in $\mathcal{G}$ all become time-varying, such as $\mathcal{E}(t)$, $\mathcal{A}(t)$ (also its elements $a_{ij}(t)$) and $\mathcal{N}_{i}(t)$. For any initial time $t_{0}\geq0$,  $\mathcal{G}(t)$, $\forall t\geq t_{0}$ is said to be {\it simultaneously structurally balanced (s.s.b.)} if $D\mathcal{A}(t)D=\left|\mathcal{A}(t)\right|$, $\forall t\geq t_{0}$ holds for some constant matrix $D\in\mathcal{D}_{n}$; and it is said to be {\it simultaneously structurally unbalanced (s.s.ub.)}, otherwise. The Laplacian matrix of $\mathcal{G}(t)$ is defined as
\begin{equation}\label{Eq1}
L(t)=\left[l_{ij}(t)\right]\in\mathbb{R}^{n\times n}~\hbox{with}~l_{ij}(t)=\left\{\aligned
&\sum_{k\in\mathcal{N}_{i}(t)}\left|a_{ik}(t)\right|,j=i\\
&-a_{ij}(t),\hspace{0.8cm}j\neq i.
\endaligned\right.
\end{equation}

Let $\mathcal{G}(t)$ switch among $M$ finite signed digraphs, i.e.,
\[\mathcal{G}(t)\in\widehat{\mathcal{G}}_{\sigma}
\triangleq\left\{\mathcal{G}_{\sigma_{p}}=\left(\mathcal{V},\mathcal{E}_{\sigma_{p}},\mathcal{A}_{\sigma_{p}}\right):p\in\mathcal{I}_{M}\right\},~~~ \forall t\geq t_{0}.\]

\noindent Furthermore, it admits some sequence $\left\{t_{k}:k\in\mathbb{Z}_{+}\right\}$ with the dwell time $t_{k+1}-t_{k}\geq\tau$ for some constant $\tau>0$ such that $\mathcal{G}(t)=\mathcal{G}_{k}$, $t\in[t_{k},t_{k+1})$ holds for some sequence of signed digraphs $\left\{\mathcal{G}_{k}=\left(\mathcal{V},\mathcal{E}_{k},\mathcal{A}_{k}\right):k\in\mathbb{Z}_{+}\right\}$ belonging to $\widehat{\mathcal{G}}_{\sigma}$. The {\it union} of a collection of signed digraphs $\left\{\mathcal{G}_{k_{1}},\mathcal{G}_{k_{2}},\cdots,\mathcal{G}_{k_{j}}\right\}$ taking in $\widehat{\mathcal{G}}_{\sigma}$ is denoted by $\bigcup_{p=1}^{j}\mathcal{G}_{k_{p}}$ whose node set is $\mathcal{V}$ and edge set is the union $\bigcup_{p=1}^{j}\mathcal{E}_{k_{p}}$ of $\mathcal{E}_{k_{p}}$ for all $p=1,2,\cdots,j$. We say that $\mathcal{G}(t)$ is {\it jointly strongly connected} if it admits some finite positive integer $h\in\mathbb{Z}$ and some resulting sequence $\left\{k_{j}:j\in\mathbb{Z}_{+}\right\}$, where $k_{0}=0$ and $0<k_{j+1}-k_{j}\leq h$, such that the union $\bigcup_{p=k_{j}}^{k_{j+1}-1}\mathcal{G}_{p}$, $\forall j\in\mathbb{Z}_{+}$ of each collection of digraphs $\left\{\mathcal{G}_{k_{j}},\mathcal{G}_{k_{j}+1},\cdots,\mathcal{G}_{k_{j+1}-1}\right\}$ is strongly connected.

\section{Separation of Cooperations and Antagonisms: A State Transition Matrix-Based Approach}\label{sec3}

Let $x(t)=\left[x_{1}(t),x_{2}(t),\cdots,x_{n}(t)\right]^{\tp}\in\mathbb{R}^{n}$ describe the opinion state. For any initial time $t_{0}\geq0$, we consider opinion dynamics in signed networks with the class of Laplacian flows given by
\begin{equation}\label{Eq3}
\dot{x}(t)=-L(t)x(t),~~~\forall t\geq t_{0}
\end{equation}

\noindent which in components is expressed as
\begin{equation*}\label{}
\dot{x}_{i}(t)=-\sum_{j\in\mathcal{N}_{i}(t)}\left|a_{ij}(t)\right|\left[x_{i}(t)-\sgn(a_{ij}(t))x_{j}(t)\right],~~~\forall i\in\mathcal{I}_{n}.
\end{equation*}

\noindent For the system (\ref{Eq3}), let $\Phi(t,t_{0})\triangleq\Phi_{-L(t)}(t,t_{0})$, $\forall t\geq t_{0}$ denote its state transition matrix in the sequel. Then the solution for the system (\ref{Eq3}) satisfies $x(t)=\Phi(t,t_{0})x_{0}$, $\forall t\geq t_{0}$ when we employ the initial state $x(t_{0})\triangleq x_{0}$.

We are interested in how to take advantage of the separation of cooperations and antagonisms such that we can develop the convergence analysis of opinion dynamics in signed networks. To this end, we resort to the state transition matrix $\Phi(t,t_{0})$ and introduce an approach based on the separation of cooperations and antagonisms to connect $\Phi(t,t_{0})$ to state transition matrices associated with conventional digraphs. It is worth pointing out that it is a general approach using no assumptions on structural balance, digon sign-symmetry or connectivity of $\mathcal{G}(t)$.

We look over the Laplacian matrix defined in (\ref{Eq1}) and arrive at
\[L(t)=\Delta_{\left|\mathcal{A}(t)\right|}-\mathcal{A}(t)\]

\noindent which can be rewritten as
\begin{equation}\label{Eq4}
\aligned
L(t)
&=\left[\Delta_{\left|\mathcal{A}(t)\right|}-\mathcal{A}^{+}(t)\right]+\left|\mathcal{A}^{-}(t)\right|\\
&=\left[\mathcal{L}_{\mathcal{A}^{+}(t)}+\Delta_{\left|\mathcal{A}^{-}(t)\right|}\right]+\left|\mathcal{A}^{-}(t)\right|,~~\forall t\geq t_{0}.
\endaligned
\end{equation}

\noindent For the additive decomposition (\ref{Eq4}), $\mathcal{L}_{\mathcal{A}^{+}(t)}+\Delta_{\left|\mathcal{A}^{-}(t)\right|}\in\mathcal{Z}_{n}$ is a diagonally dominant matrix that shares the same non-diagonal entries as the nonpositive matrix $-\mathcal{A}^{+}(t)\leq0$ and $\left|\mathcal{A}^{-}(t)\right|\geq0$ is a nonnegative matrix. With these two properties, if we insert (\ref{Eq4}) into (\ref{Eq3}), then we consider the solving of linear systems and can propose useful properties in the lemma below.

\begin{lem}\label{Lem8}
For any $t\geq t_{0}$, $\Phi_{-\mathcal{L}_{\mathcal{A}^{+}(t)}-\Delta_{\left|\mathcal{A}^{-}(t)\right|}}(t,t_{0})\geq0$ holds and is generally a substochastic matrix such that
\begin{equation}\label{Eq6}
\aligned
\Phi(t,t_{0})
&=\Phi_{-\mathcal{L}_{\mathcal{A}^{+}(t)}-\Delta_{\left|\mathcal{A}^{-}(t)\right|}}(t,t_{0})\\
&~~~-\int_{t_{0}}^{t}\Phi_{-\mathcal{L}_{\mathcal{A}^{+}(t)}-\Delta_{\left|\mathcal{A}^{-}(t)\right|}}(t,\theta)
\left|\mathcal{A}^{-}(\theta)\right|\Phi(\theta,t_{0})d\theta.
\endaligned
\end{equation}
\end{lem}

\begin{IEEEproof}
See Appendix \ref{ap1}.
\end{IEEEproof}

For the trivial case if $\mathcal{A}^{-}(t)\equiv0$, $\forall t\geq t_{0}$ holds, we have $\mathcal{A}(t)\equiv\mathcal{A}^{+}(t)\geq0$, and consequently $\Phi_{-\mathcal{L}_{\mathcal{A}^{+}(t)}-\Delta_{\left|\mathcal{A}^{-}(t)\right|}}(t,t_{0})=\Phi_{-\mathcal{L}_{\mathcal{A}(t)}}(t,t_{0})$, $\forall t\geq t_{0}$ becomes a stochastic matrix. This, together with (\ref{Eq6}), yields that $\Phi(t,t_{0})=\Phi_{-\mathcal{L}_{\mathcal{A}(t)}}(t,t_{0})$, $\forall t\geq t_{0}$ is a stochastic matrix. It is actually implied that our developments include the counterpart results of conventional networks as a marginal case. Otherwise, if $\mathcal{A}^{-}(t)\not\equiv0$, $\forall t\geq t_{0}$ holds, then $\Phi_{-\mathcal{L}_{\mathcal{A}^{+}(t)}-\Delta_{\left|\mathcal{A}^{-}(t)\right|}}(t,t_{0})\geq0$ is generally a substochastic matrix associated with a conventional digraph that involves a nonnegative adjacency matrix. Despite of this fact, $\Phi(t,t_{0})$ is embedded in the relation (\ref{Eq6}), from which there can not be seen a clear relationship between $\Phi(t,t_{0})$ and $\Phi_{-\mathcal{L}_{\mathcal{A}^{+}(t)}-\Delta_{\left|\mathcal{A}^{-}(t)\right|}}(t,t_{0})$. To overcome this drawback, we obtain a further expression for the relationship between them.

\begin{lem}\label{Lem6}
For any $t\geq t_{0}$, $\Phi(t,t_{0})$ has a series expression of
\begin{equation}\label{Eq5}
\aligned
\Phi(t,t_{0})
&=\Phi_{-\mathcal{L}_{\mathcal{A}^{+}(t)}-\Delta_{\left|\mathcal{A}^{-}(t)\right|}}(t,t_{0})\\
&~~~+\sum_{k=1}^{\infty}(-1)^{k}\int_{t_{0}}^{t}\Phi_{-\mathcal{L}_{\mathcal{A}^{+}(t)}-\Delta_{\left|\mathcal{A}^{-}(t)\right|}}(t,\theta_{1})\left|\mathcal{A}^{-}(\theta_{1})\right|\\
&~~~\times\int_{t_{0}}^{\theta_{1}}
\Phi_{-\mathcal{L}_{\mathcal{A}^{+}(t)}-\Delta_{\left|\mathcal{A}^{-}(t)\right|}}(\theta_{1},\theta_{2})\left|\mathcal{A}^{-}(\theta_{2})\right|\cdots\\
&~~~\times\int_{t_{0}}^{\theta_{k-1}}
\Phi_{-\mathcal{L}_{\mathcal{A}^{+}(t)}-\Delta_{\left|\mathcal{A}^{-}(t)\right|}}(\theta_{k-1},\theta_{k})\left|\mathcal{A}^{-}(\theta_{k})\right|\\
&~~~\times\Phi_{-\mathcal{L}_{\mathcal{A}^{+}(t)}-\Delta_{\left|\mathcal{A}^{-}(t)\right|}}(\theta_{k},t_{0})d\theta_{k}\cdots d\theta_{2}d\theta_{1}
\endaligned
\end{equation}

\noindent where (and afterwards) $\theta_{0}\triangleq t$.
\end{lem}

\begin{IEEEproof}
By Lemma \ref{Lem8}, if we further exploit the relationship (\ref{Eq6}), then we can arrive at the series expression (\ref{Eq5}) for $\Phi(t,t_{0})$ by following the same steps as the derivation of the Peano-Baker series expression for the state transition matrix of linear systems (see, e.g., \cite[eq. (12), p. 44]{r:96}), which is omitted here for simplicity.
\end{IEEEproof}

The series expression (\ref{Eq5}) may help provide a possible way to exploit convergence properties of $\Phi(t,t_{0})$ for signed networks based on the nonnegative matrix theory of \cite{hj:85}. In other words, we may incorporate Lemma \ref{Lem6} to bridge a relationship between the analyses of signed networks and of conventional networks. To make this point clearer to follow, we establish the following theorem to present an additive decomposition of $\Phi(t,t_{0})$ based on two nonnegative matrices.

\begin{thm}\label{Thm4}
For any $t\geq t_{0}$, $\Phi(t,t_{0})$ can be decomposed into
\begin{equation}\label{Eq23}
\Phi(t,t_{0})=\Phi_{\mathrm{even}}(t,t_{0})-\Phi_{\mathrm{odd}}(t,t_{0})
\end{equation}

\noindent where $\Phi_{\mathrm{even}}(t,t_{0})\geq0$, $\forall t\geq t_{0}$ and $\Phi_{\mathrm{odd}}(t,t_{0})\geq0$, $\forall t\geq t_{0}$ are two nonnegative matrices given by
\begin{equation}\label{Eq7}
\aligned
\Phi_{\mathrm{even}}(t,t_{0})
&=\Phi_{-\mathcal{L}_{\mathcal{A}^{+}(t)}-\Delta_{\left|\mathcal{A}^{-}(t)\right|}}(t,t_{0})\\
&~~~+\sum_{k=1}^{\infty}\int_{t_{0}}^{t}\Phi_{-\mathcal{L}_{\mathcal{A}^{+}(t)}-\Delta_{\left|\mathcal{A}^{-}(t)\right|}}(t,\theta_{1})\left|\mathcal{A}^{-}(\theta_{1})\right|\\
&~~~\times\int_{t_{0}}^{\theta_{1}}
\Phi_{-\mathcal{L}_{\mathcal{A}^{+}(t)}-\Delta_{\left|\mathcal{A}^{-}(t)\right|}}(\theta_{1},\theta_{2})\left|\mathcal{A}^{-}(\theta_{2})\right|\cdots\\
&~~~\times\int_{t_{0}}^{\theta_{2k-1}}
\Phi_{-\mathcal{L}_{\mathcal{A}^{+}(t)}-\Delta_{\left|\mathcal{A}^{-}(t)\right|}}(\theta_{2k-1},\theta_{2k})\left|\mathcal{A}^{-}(\theta_{2k})\right|\\
&~~~\times\Phi_{-\mathcal{L}_{\mathcal{A}^{+}(t)}-\Delta_{\left|\mathcal{A}^{-}(t)\right|}}(\theta_{2k},t_{0})d\theta_{2k}\cdots d\theta_{2}d\theta_{1},~~~\forall t\geq t_{0}
\endaligned
\end{equation}

\noindent and
\begin{equation}\label{Eq8}
\aligned
\Phi_{\mathrm{odd}}(t,t_{0})
&=\sum_{k=0}^{\infty}\int_{t_{0}}^{t}\Phi_{-\mathcal{L}_{\mathcal{A}^{+}(t)}-\Delta_{\left|\mathcal{A}^{-}(t)\right|}}(t,\theta_{1})\left|\mathcal{A}^{-}(\theta_{1})\right|\\
&~~~\times\int_{t_{0}}^{\theta_{1}}
\Phi_{-\mathcal{L}_{\mathcal{A}^{+}(t)}-\Delta_{\left|\mathcal{A}^{-}(t)\right|}}(\theta_{1},\theta_{2})\left|\mathcal{A}^{-}(\theta_{2})\right|\cdots\\
&~~~\times\int_{t_{0}}^{\theta_{2k}}
\Phi_{-\mathcal{L}_{\mathcal{A}^{+}(t)}-\Delta_{\left|\mathcal{A}^{-}(t)\right|}}(\theta_{2k},\theta_{2k+1})\left|\mathcal{A}^{-}(\theta_{2k+1})\right|\\
&~~~\times\Phi_{-\mathcal{L}_{\mathcal{A}^{+}(t)}-\Delta_{\left|\mathcal{A}^{-}(t)\right|}}(\theta_{2k+1},t_{0})d\theta_{2k+1}\cdots d\theta_{2}d\theta_{1},~~~\forall t\geq t_{0}.
\endaligned
\end{equation}
\end{thm}

\begin{IEEEproof}
We note a fact that
\[\frac{\Ds d\Phi_{-\mathcal{L}_{\mathcal{A}^{+}(t)}-\Delta_{\left|\mathcal{A}^{-}(t)\right|}}(t,\theta_{1})}{\Ds dt}
=-\left[\mathcal{L}_{\mathcal{A}^{+}(t)}+\Delta_{\left|\mathcal{A}^{-}(t)\right|}\right]
\Phi_{-\mathcal{L}_{\mathcal{A}^{+}(t)}-\Delta_{\left|\mathcal{A}^{-}(t)\right|}}(t,\theta_{1}).
\]

\noindent Then using (\ref{Eq7}) and (\ref{Eq8}), we can verify
\begin{equation}\label{Eq9}
\aligned
\frac{\Ds d\Phi_{\mathrm{even}}(t,t_{0})}{\Ds dt}
&=-\left[\mathcal{L}_{\mathcal{A}^{+}(t)}+\Delta_{\left|\mathcal{A}^{-}(t)\right|}\right]\Phi_{\mathrm{even}}(t,t_{0})\\
&~~~+\left|\mathcal{A}^{-}(t)\right|\Phi_{\mathrm{odd}}(t,t_{0}),~~~\forall t\geq t_{0}
\endaligned
\end{equation}

\noindent and
\begin{equation}\label{Eq10}
\aligned
\frac{\Ds d\Phi_{\mathrm{odd}}(t,t_{0})}{\Ds d t}
&=-\left[\mathcal{L}_{\mathcal{A}^{+}(t)}+\Delta_{\left|\mathcal{A}^{-}(t)\right|}\right]\Phi_{\mathrm{odd}}(t,t_{0})\\
&~~~+\left|\mathcal{A}^{-}(t)\right|\Phi_{\mathrm{even}}(t,t_{0}),~~~\forall t\geq t_{0}.
\endaligned
\end{equation}

\noindent Consequently, we consider (\ref{Eq4}), (\ref{Eq9}) and (\ref{Eq10}) to obtain
\begin{equation}\label{Eq24}
\aligned
\frac{\Ds d\left[\Phi_{\mathrm{even}}(t,t_{0})-\Phi_{\mathrm{odd}}(t,t_{0})\right]}{\Ds dt}
&=-\left[\mathcal{L}_{\mathcal{A}^{+}(t)}+\Delta_{\left|\mathcal{A}^{-}(t)\right|}+\left|\mathcal{A}^{-}(t)\right|\right]
\left[\Phi_{\mathrm{even}}(t,t_{0})-\Phi_{\mathrm{odd}}(t,t_{0})\right]\\
&=-L(t)\left[\Phi_{\mathrm{even}}(t,t_{0})-\Phi_{\mathrm{odd}}(t,t_{0})\right],~~~\forall t\geq t_{0}.
\endaligned
\end{equation}

\noindent We can see also from (\ref{Eq7}) and (\ref{Eq8}) that
\begin{equation}\label{Eq25}
\aligned
\Phi_{\mathrm{even}}(t_{0},t_{0})&=\Phi_{-\mathcal{L}_{\mathcal{A}^{+}(t)}-\Delta_{\left|\mathcal{A}^{-}(t)\right|}}(t_{0},t_{0})=I\\
\Phi_{\mathrm{odd}}(t_{0},t_{0})&=0.
\endaligned
\end{equation}

\noindent For the initial condition of (\ref{Eq24}), we can deduce from (\ref{Eq25}) that $\Phi_{\mathrm{even}}(t_{0},t_{0})-\Phi_{\mathrm{odd}}(t_{0},t_{0})=I$. That is, $\Phi_{\mathrm{even}}(t,t_{0})-\Phi_{\mathrm{odd}}(t,t_{0})$ has the same properties as $\Phi(t,t_{0})$. Because the state transition matrix of the linear system (\ref{Eq3}) is unique (see also \cite{r:96}), it is immediate to obtain (\ref{Eq23}). Owing to $\Phi_{-\mathcal{L}_{\mathcal{A}^{+}(t)}-\Delta_{\left|\mathcal{A}^{-}(t)\right|}}(t,t_{0})\geq0$, $\forall t\geq t_{0}$ from Lemma \ref{Lem8}, $\Phi_{\mathrm{even}}(t,t_{0})\geq0$ and $\Phi_{\mathrm{odd}}(t,t_{0})\geq0$ for all $t\geq t_{0}$ can be verified based on (\ref{Eq7}) and (\ref{Eq8}), respectively.
\end{IEEEproof}

By comparing (\ref{Eq7}) and (\ref{Eq8}) with (\ref{Eq5}), we notice that $\Phi_{\mathrm{even}}(t,t_{0})$ and $\Phi_{\mathrm{odd}}(t,t_{0})$ are in fact composed of the even and (negative) odd terms of the series expression (\ref{Eq5}), respectively. Moreover, this composition style motivates us to naturally ask a question: what is the sum of $\Phi_{\mathrm{even}}(t,t_{0})$ and $\Phi_{\mathrm{odd}}(t,t_{0})$? Are there further properties for nonnegative matrices $\Phi_{\mathrm{even}}(t,t_{0})$ and $\Phi_{\mathrm{odd}}(t,t_{0})$? To answer the two questions, we resort to $\Phi_{-\mathcal{L}_{\left|\mathcal{A}(t)\right|}}(t,t_{0})$. It is worth noting that $\Phi_{-\mathcal{L}_{\left|\mathcal{A}(t)\right|}}(t,t_{0})\geq0$, $\forall t\geq t_{0}$ is a stochastic matrix under the conventional digraph associated with $\left|\mathcal{A}(t)\right|\geq0$ (see, e.g., \cite{rb:05}). Next, we formally present a theorem to bridge a clear relation between two state transition matrices $\Phi(t,t_{0})$ and $\Phi_{-\mathcal{L}_{\left|\mathcal{A}(t)\right|}}(t,t_{0})$.

\begin{thm}\label{Thm2}
The stochastic matrix $\Phi_{-\mathcal{L}_{\left|\mathcal{A}(t)\right|}}(t,t_{0})\geq0$, $\forall t\geq t_{0}$ can be expressed as
\begin{equation}\label{Eq26}
\Phi_{-\mathcal{L}_{\left|\mathcal{A}(t)\right|}}(t,t_{0})=\Phi_{\mathrm{even}}(t,t_{0})+\Phi_{\mathrm{odd}}(t,t_{0}).
\end{equation}

\noindent Moreover, $\Phi(t,t_{0})$ can be bounded by $\Phi_{-\mathcal{L}_{\left|\mathcal{A}(t)\right|}}(t,t_{0})$ such that
\begin{equation}\label{Eq27}
\left|\Phi(t,t_{0})\right|\leq\Phi_{-\mathcal{L}_{\left|\mathcal{A}(t)\right|}}(t,t_{0}),~~~\forall t\geq t_{0}
\end{equation}

\noindent or equivalently, $-\Phi_{-\mathcal{L}_{\left|\mathcal{A}(t)\right|}}(t,t_{0})\leq\Phi(t,t_{0})\leq\Phi_{-\mathcal{L}_{\left|\mathcal{A}(t)\right|}}(t,t_{0})$, $\forall t\geq t_{0}$ holds.
\end{thm}

\begin{IEEEproof}
In the same way as performed in (\ref{Eq24}), we can use (\ref{Eq9}) and (\ref{Eq10}) to deduce
\[\aligned
\frac{\Ds d\left[\Phi_{\mathrm{even}}(t,t_{0})+\Phi_{\mathrm{odd}}(t,t_{0})\right]}{\Ds dt}
&=-\left[\mathcal{L}_{\mathcal{A}^{+}(t)}+\Delta_{\left|\mathcal{A}^{-}(t)\right|}-\left|\mathcal{A}^{-}(t)\right|\right]
\left[\Phi_{\mathrm{even}}(t,t_{0})+\Phi_{\mathrm{odd}}(t,t_{0})\right]\\
&=-\mathcal{L}_{\left|\mathcal{A}(t)\right|}\left[\Phi_{\mathrm{even}}(t,t_{0})+\Phi_{\mathrm{odd}}(t,t_{0})\right],~~~\forall t\geq t_{0}
\endaligned
\]

\noindent for which we can easily get the initial condition $\Phi_{\mathrm{even}}(t_{0},t_{0})+\Phi_{\mathrm{odd}}(t_{0},t_{0})=I$ from (\ref{Eq25}). Hence, both $\Phi_{\mathrm{even}}(t,t_{0})+\Phi_{\mathrm{odd}}(t,t_{0})$ and $\Phi_{-\mathcal{L}_{\left|\mathcal{A}(t)\right|}}(t,t_{0})$ represent the state transition matrix of the linear system with the state matrix $-\mathcal{L}_{\left|\mathcal{A}(t)\right|}$, and consequently (\ref{Eq26}) holds for the same reason as used in the derivation of (\ref{Eq23}). Again by considering (\ref{Eq23}) and (\ref{Eq26}), we employ \cite[eq. (8.1.3)]{hj:85} to validate
\[\aligned
\left|\Phi(t,t_{0})\right|
&=\left|\Phi_{\mathrm{even}}(t,t_{0})-\Phi_{\mathrm{odd}}(t,t_{0})\right|\\
&\leq\Phi_{\mathrm{even}}(t,t_{0})+\Phi_{\mathrm{odd}}(t,t_{0})\\
&=\Phi_{-\mathcal{L}_{\left|\mathcal{A}(t)\right|}}(t,t_{0}),~~\forall t\geq t_{0}\geq0
\endaligned\]

\noindent i.e., (\ref{Eq27}) holds.
\end{IEEEproof}

\begin{rem}\label{Rem4}
In Theorems \ref{Thm4} and \ref{Thm2}, we show relations between both state transition matrices $\Phi(t,t_{0})$ and $\Phi_{-\mathcal{L}_{\left|\mathcal{A}(t)\right|}}(t,t_{0})$ which represent the distributed Laplacian flows under signed digraphs and conventional digraphs, respectively. This indicates that the relations (\ref{Eq23}), (\ref{Eq26}) and (\ref{Eq27}) may contribute to developing a new approach capable of bridging the gap between the convergence analysis of signed networks and of conventional networks. It is worth noticing that they require no hypotheses on connectivity, structural balance, or digon sign-symmetry of signed digraphs, but benefit from the introduction of the series expressions (\ref{Eq5}), (\ref{Eq7}) and (\ref{Eq8}). Moreover, the three series all converge absolutely, which can be validated based on the relations (\ref{Eq23}), (\ref{Eq26}) and (\ref{Eq27}) together with the absolute convergence of the Peano-Baker series of $\Phi_{-\mathcal{L}_{\left|\mathcal{A}(t)\right|}}(t,t_{0})$ (see also \cite{r:96}).
\end{rem}

As a direct application of Theorems \ref{Thm4} and \ref{Thm2}, we can develop further properties of $\Phi_{\mathrm{even}}(t,t_{0})$, $\Phi_{\mathrm{odd}}(t,t_{0})$ and $\Phi(t,t_{0})$.

\begin{cor}\label{Cor1}
For any $t\geq t_{0}$, both $\Phi_{\mathrm{even}}(t,t_{0})$ and $\Phi_{\mathrm{odd}}(t,t_{0})$ are substochastic matrices, and $\Phi(t,t_{0})$ satisfies
\begin{equation}\label{Eq11}
\left\|\Phi(t,t_{0})\right\|_{\infty}\leq1,~~~\forall t\geq t_{0}.
\end{equation}
\end{cor}

\begin{IEEEproof}
Note that $\Phi_{\mathrm{even}}(t,t_{0})\geq0$, $\forall t\geq t_{0}$ and $\Phi_{\mathrm{odd}}(t,t_{0})\geq0$, $\forall t\geq t_{0}$ are derived in Theorem \ref{Thm4} and $\Phi_{-\mathcal{L}_{\left|\mathcal{A}(t)\right|}}(t,t_{0})1_{n}=1_{n}$, $\forall t\geq t_{0}$ is satisfied for the stochastic matrix $\Phi_{-\mathcal{L}_{\left|\mathcal{A}(t)\right|}}(t,t_{0})$. By these facts, we can apply (\ref{Eq26}) to obtain
\[\aligned
\Phi_{\mathrm{even}}(t,t_{0})1_{n}
&=\Phi_{-\mathcal{L}_{\left|\mathcal{A}(t)\right|}}(t,t_{0})1_{n}-\Phi_{\mathrm{odd}}(t,t_{0})1_{n}\\
&\leq\Phi_{-\mathcal{L}_{\left|\mathcal{A}(t)\right|}}(t,t_{0})1_{n}\\
&=1_{n},~~~\forall t\geq t_{0}
\endaligned
\]

\noindent which guarantees that $\Phi_{\mathrm{even}}(t,t_{0})\geq0$, $\forall t\geq t_{0}$ is a substochastic matrix. In the same way, we can deduce also that $\Phi_{\mathrm{odd}}(t,t_{0})\geq0$, $\forall t\geq t_{0}$ is a substochastic matrix. By noticing $\left\|\Phi(t,t_{0})\right\|_{\infty}=\big\|\left|\Phi(t,t_{0})\right|\big\|_{\infty}$ and $\left\|\Phi_{-\mathcal{L}_{\left|\mathcal{A}(t)\right|}}(t,t_{0})\right\|_{\infty}=1$, we can develop that (\ref{Eq11}) is an immediate consequence of the relation (\ref{Eq27}).
\end{IEEEproof}

\begin{rem}\label{Rem1}
By Corollary \ref{Cor1}, the relations (\ref{Eq23}) and (\ref{Eq26}) further reveal that the state transition matrix $\Phi(t,t_{0})$ under the signed digraph of an adjacency matrix $\mathcal{A}(t)$ is the difference between two substochastic matrices whose sum is the stochastic matrix $\Phi_{-\mathcal{L}_{\left|\mathcal{A}(t)\right|}}(t,t_{0})$, i.e., the state transition matrix under a resulting conventional digraph of the adjacency matrix $\left|\mathcal{A}(t)\right|\geq0$. If we resort to \cite[Definition 6.1]{r:96}, (\ref{Eq11}) ensures that the system (\ref{Eq3}) is uniformly stable, and consequently $\left\|x(t)\right\|_{\infty}\leq\left\|x_{0}\right\|_{\infty}$, $\forall t\geq t_{0}$ is bounded. This gives a fundamental premise for our considered convergence problems on signed networks. For example, (\ref{Eq11}) ensures the stability of the system (\ref{Eq3}) actually to be uniformly asymptotically/exponentially stable by noticing \cite[Definition 6.12 and Theorem 6.13]{r:96}.
\end{rem}

In Theorem \ref{Thm2}, the relation (\ref{Eq27}) suggests a direct estimation which may make it possible to achieve convergence of signed networks by addressing a counterpart problem of conventional networks. In fact, Theorems \ref{Thm4} and \ref{Thm2} can be exploited to obtain a convergence analysis approach for signed networks such that the existing convergence results of conventional networks can be incorporated after a state transition matrix-based analysis is performed. This is thus different from the existing approaches employed in, e.g., \cite{a:13,vm:14,hz:14,m:17,h:14,xcj:16,pmc:16}, which is revealed more clearly in the following theorem.

\begin{thm}\label{Thm1}
Let
\begin{equation}\label{Eq28}
\Psi(t,t_{0})
=\begin{bmatrix}\Phi_{\mathrm{even}}(t,t_{0})&\Phi_{\mathrm{odd}}(t,t_{0})\\\Phi_{\mathrm{odd}}(t,t_{0})&\Phi_{\mathrm{even}}(t,t_{0})\end{bmatrix},~~~\forall t\geq t_{0}.
\end{equation}

\noindent Then $\Psi(t,t_{0})=\Phi_{-\mathcal{L}_{\mathbf{A}(t)}}(t,t_{0})$ defines the state transition matrix of a linear system whose state matrix is expressed by $-\mathcal{L}_{\mathbf{A}(t)}$, where $\mathcal{L}_{\mathbf{A}(t)}$ is exactly the Laplacian matrix associated with a nonnegative matrix $\mathbf{A}(t)\geq0$ such that
\[
\mathbf{A}(t)
=\begin{bmatrix}\mathcal{A}^{+}(t)&\left|\mathcal{A}^{-}(t)\right|\\
\left|\mathcal{A}^{-}(t)\right|&\mathcal{A}^{+}(t)\end{bmatrix}\geq0,~~\forall t\geq t_{0}.
\]
\end{thm}

\begin{IEEEproof}
With $\Psi(t,t_{0})$ in (\ref{Eq28}), we revisit (\ref{Eq9}) and (\ref{Eq10}) and can show that they can be reformulated in a compact form of
\begin{equation}\label{Eq12}
\aligned
\frac{\Ds d\Psi(t,t_{0})}{\Ds dt}&=-\mathcal{L}_{\mathbf{A}(t)}\Psi(t,t_{0}),~~\forall t\geq t_{0}
\endaligned
\end{equation}

\noindent where $\mathcal{L}_{\mathbf{A}(t)}$ is given by
\begin{equation}\label{Eq13}
\aligned
\mathcal{L}_{\mathbf{A}(t)}
&=\begin{bmatrix}\mathcal{L}_{\mathcal{A}^{+}(t)}+\Delta_{\left|\mathcal{A}^{-}(t)\right|}&-\left|\mathcal{A}^{-}(t)\right|\\
-\left|\mathcal{A}^{-}(t)\right|&\mathcal{L}_{\mathcal{A}^{+}(t)}+\Delta_{\left|\mathcal{A}^{-}(t)\right|}\end{bmatrix}.
\endaligned
\end{equation}

\noindent We can see from (\ref{Eq25}) that $\Psi(t_{0},t_{0})=I$ holds at the initial time $t_{0}$. Hence, the uniqueness of solution to the matrix differential equation (\ref{Eq12}) leads to $\Psi(t,t_{0})=\Phi_{-\mathcal{L}_{\mathbf{A}(t)}}(t,t_{0})$. From (\ref{Eq13}), it is clear to see that $\mathcal{L}_{\mathbf{A}(t)}$ defines the Laplacian matrix associated with $\mathbf{A}(t)\geq0$.
\end{IEEEproof}

From (\ref{Eq23}) and (\ref{Eq28}), we can solve the convergence problem of $\Phi(t,t_{0})$ if we can instead achieve the convergence of $\Psi(t,t_{0})$. If we further consider (\ref{Eq26}), then $\lim_{t\to\infty}\Psi(t,t_{0})$ is accomplished if and only if $\lim_{t\to\infty}\Phi(t,t_{0})$ and $\lim_{t\to\infty}\Phi_{-\mathcal{L}_{\left|\mathcal{A}(t)\right|}}(t,t_{0})$ can be both achieved. These properties provide a feasible approach to addressing convergence problems on signed networks through coping with the resulting convergence problems on counterpart conventional networks. To this end, we present some notations about the conventional digraph associated with the nonnegative matrix $\mathbf{A}(t)\geq0$.

Since $\mathbf{A}(t)\geq0$ defines an $2n\times2n$ nonnegative matrix that is induced from $\mathcal{A}(t)\in\mathbb{R}^{n\times n}$, we can correspondingly introduce a conventional digraph $\mathbf{G}(t)=\left(\mathbf{V},\mathbf{E}(t),\mathbf{A}(t)\right)$ which is induced from the signed digraph $\mathcal{G}(t)$ but not signed any longer. Due to $\mathcal{G}(t)\in\widehat{\mathcal{G}}_{\sigma}$, $\forall t\geq t_{0}$, we accordingly have $\mathbf{G}(t)\in\widehat{\mathbf{G}}_{\sigma}$, $\forall t\geq t_{0}$ for some $\widehat{\mathbf{G}}_{\sigma}=\left\{\mathbf{G}_{\sigma_{1}},\mathbf{G}_{\sigma_{2}},\cdots,\mathbf{G}_{\sigma_{M}}\right\}$, where $\mathbf{G}_{\sigma_{p}}$, $\forall p\in\mathcal{I}_{M}$ is induced from $\mathcal{G}_{\sigma_{p}}$ in the same way as $\mathbf{G}(t)$ is induced from $\mathcal{G}(t)$. Similarly, let us denote $\mathbf{G}_{k}$, $\forall k\in\mathbb{Z}_{+}$ which is induced from $\mathcal{G}_{k}$. Then by considering the time sequence $\left\{t_{k}:k\in\mathbb{Z}_{+}\right\}$, we have $\mathbf{G}(t)=\mathbf{G}_{k}$, $\forall t\in[t_{k},t_{k+1})$ and $\mathbf{G}_{k}\in\widehat{\mathbf{G}}_{\sigma}$, $\forall k\in\mathbb{Z}_{+}$. It is easy to observe that the signed digraph $\mathcal{G}(t)$ can be closely tied to the conventional digraph $\mathbf{G}(t)$.

\begin{rem}\label{Rem5}
The above analysis exploited by the formulation (\ref{Eq28}) is similar to that derived by the lifting approach (see, e.g., \cite{h:14,xcj:16}). However, different from these existing results, we contribute mainly to the continuous-time model for opinion dynamics and introduce a state transition matrix-based analysis approach. Particularly, unlike the conclusions that are made for the lifting approach (see \cite[p. 2122]{h:14}), the analysis performed in this paper does not need the digon sign-symmetry of signed digraphs, i.e., it admits the cases $a_{ij}(t)a_{ji}(t^{\prime})<0$, $\forall t,t^{\prime}\geq t_{0}$ for some $i,j\in\mathcal{I}_{n}$.
\end{rem}

\section{State Transition Matrix-Based Convergence Analysis of Opinion Dynamics}\label{sec4}

In this section, we study under what conditions we can reach the convergence analysis of the state transition matrix $\Phi(t,t_{0})$. We then establish a convergence analysis approach for opinion dynamics in signed networks based on taking advantage of the well-developed convergence results for conventional networks.

We are concerned with asymptotic behaviors for the network flows described by (\ref{Eq3}). If it admits some $c\in\mathbb{R}$ and some $D\in\mathcal{D}_{n}$ such that $\lim_{t\to\infty}x(t)=c\left(D1_{n}\right)$, then we say that the system (\ref{Eq3}) achieves {\it bipartite consensus} or {\it polarizes}. If $\lim_{t\to\infty}x(t)=0$ is accomplished, then the system (\ref{Eq3}) is said to be {\it stable} or to {\it neutralize}. These two asymptotic problems of polarization and neutralization will be addressed in this paper and be connected to the s.s.b. and s.s.ub. properties of $\mathcal{G}(t)$, respectively. Toward this end, we make an assumption on the switching of $\mathcal{G}(t)$.

\begin{assum}\label{Assum1}
For any $p\in\mathcal{I}_{M}$ and any $\hat{t}\geq t_{0}$, there follows $\left\{t:\mathcal{G}(t)=\mathcal{G}_{\sigma_{p}}\right\}\bigcap\left[\hat{t},\hat{t}+T\right]\neq\O$ for some constant $T>0$.
\end{assum}

With Assumption \ref{Assum1}, each alternative selection $\mathcal{G}_{\sigma_{p}}$, $\forall p\in\mathcal{I}_{M}$ plays an important part in the switching process of $\mathcal{G}(t)$, $\forall t\geq t_{0}$. This provides a basic guarantee such that the simultaneous structure balance effectively works in the asymptotic behaviors of the system (\ref{Eq3}). In particular, Assumption \ref{Assum1} naturally holds for the fixed topology case when $\mathcal{G}(t)\equiv\mathcal{G}$, $\forall t\geq t_{0}$.

A helpful property embedded in Assumption \ref{Assum1} is presented in the following lemma.

\begin{lem}\label{Lem1}
If Assumption \ref{Assum1} holds, then there exists some sequence $\left\{k_{j}:j\in\mathbb{Z}_{+}\right\}$ with $k_{0}=0$ and $0<k_{j+1}-k_{j}\leq h$ for some finite positive integer $h\in\mathbb{Z}$ such that for each collection of digraphs $\left\{\mathcal{G}_{k_{j}},\mathcal{G}_{k_{j}+1},\cdots,\mathcal{G}_{k_{j+1}-1}\right\}$, $\bigcup_{p=k_{j}}^{k_{j+1}-1}\mathcal{G}_{p}=\bigcup_{p=1}^{M}\mathcal{G}_{\sigma_{p}}$, $\forall j\in\mathbb{Z}_{+}$ holds. Furthermore, if the union $\bigcup_{p=1}^{M}\mathcal{G}_{\sigma_{p}}$ is strongly connected, then $\mathcal{G}(t)$ is jointly strongly connected.
\end{lem}

\begin{IEEEproof}
Let us particularly take $h=\mathrm{int}\left[T/\tau\right]+1$. Then for any finite positive integer $\hat{h}\geq h$, we can obtain some sequence $\left\{k_{j}:j\in\mathbb{Z}_{+}\right\}$ with $k_{0}=0$ and $h\leq k_{j+1}-k_{j}\leq\hat{h}$, $\forall j\in\mathbb{Z}_{+}$. This leads to $\sum_{p=k_{j}}^{k_{j+1}-1}\tau_{p}\geq h\tau>T$, $\forall j\in\mathbb{Z}_{+}$, with which we have
\begin{equation}\label{Eq22}
\left[t_{k_{j}},t_{k_{j+1}}\right)
=\left[t_{k_{j}},t_{k_{j}}+\sum_{p=k_{j}}^{k_{j+1}-1}\tau_{p}\right)
\supset\left[t_{k_{j}},t_{k_{j}}+T\right],~~\forall j\in\mathbb{Z}_{+}.
\end{equation}

\noindent As a consequence of (\ref{Eq22}), we can derive by Assumption \ref{Assum1} that $\bigcup_{p=k_{j}}^{k_{j+1}-1}\mathcal{G}_{p}\supseteq\bigcup_{t\in\left[t_{k_{j}},t_{k_{j}}+T\right]}\mathcal{G}(t)\supseteq\bigcup_{p=1}^{M}\mathcal{G}_{\sigma_{p}}$, $\forall j\in\mathbb{Z}_{+}$. Because of $\bigcup_{k=0}^{\infty}\mathcal{G}_{k}=\bigcup_{t\geq t_{0}}\mathcal{G}(t)=\bigcup_{p=1}^{M}\mathcal{G}_{\sigma_{p}}$, it follows $\bigcup_{p=k_{j}}^{k_{j+1}-1}\mathcal{G}_{p}=\bigcup_{p=1}^{M}\mathcal{G}_{\sigma_{p}}$, $\forall j\in\mathbb{Z}_{+}$. With this fact, the joint strong connectivity is immediate from the strong connectivity of $\bigcup_{p=1}^{M}\mathcal{G}_{\sigma_{p}}$.
\end{IEEEproof}

Next, we take advantage of the state transition matrix-based results in Theorems \ref{Thm4}-\ref{Thm1} to exploit the convergence analysis of $\Phi(t,t_{0})$. To proceed with this discussion, we give a preliminary lemma for the convergence of $\Phi_{-\mathcal{L}_{\left|\mathcal{A}(t)\right|}}(t,t_{0})$.

\begin{lem}\label{Lem2}
Let Assumption \ref{Assum1} hold and the union $\bigcup_{p=1}^{M}\mathcal{G}_{\sigma_{p}}$ be strongly connected. Then for any $t_{0}\geq0$, there exists some nonnegative vector $\nu(t_{0})\geq0$ satisfying $\nu^{\tp}(t_{0})1_{n}=1$ such that $\lim_{t\to\infty}\Phi_{-\mathcal{L}_{\left|\mathcal{A}(t)\right|}}(t,t_{0})=1_{n}\nu^{\tp}(t_{0})$ holds, together with its limit being approached exponentially fast.
\end{lem}

\begin{IEEEproof}
Note that the signed digraph $\mathcal{G}(t)$ and the conventional digraph $\mathcal{G}(\left|\mathcal{A}(t)\right|)$ associated with $\left|\mathcal{A}(t)\right|\geq0$ share the same connectivity. From Lemma \ref{Lem1}, it follows that Assumption \ref{Assum1} and the strong connectivity of $\bigcup_{p=1}^{M}\mathcal{G}_{\sigma_{p}}$ can ensure the joint strong connectivity of $\mathcal{G}(\left|\mathcal{A}(t)\right|)$. Consequently, the exponential convergence result for the stochastic matrix $\Phi_{-\mathcal{L}_{\left|\mathcal{A}(t)\right|}}(t,t_{0})$ can be directly obtained from the existing literatures (see, e.g., \cite{rb:05,mj:16}).
\end{IEEEproof}

With Lemma \ref{Lem2} and based on Theorems \ref{Thm4} and \ref{Thm1}, we establish a theorem for the convergence of $\Phi(t,t_{0})$ first in the case when it is associated with a s.s.b. signed digraph $\mathcal{G}(t)$.

\begin{thm}\label{Thm3}
Let Assumption \ref{Assum1} hold and the union $\bigcup_{p=1}^{M}\mathcal{G}_{\sigma_{p}}$ be strongly connected. If $\mathcal{G}(t)$ is s.s.b., then $\lim_{t\to\infty}\Phi(t,t_{0})=D1_{n}\nu^{\tp}(t_{0})D$ holds with its limit being approached exponentially fast, where $D\in\mathcal{D}_{n}$ is such that $D\mathcal{A}(t)D=\left|\mathcal{A}(t)\right|$, $\forall t\geq t_{0}$.
\end{thm}

\begin{IEEEproof}
When $\mathcal{G}(t)$ is s.s.b., we have $D\mathcal{A}(t)D=\left|\mathcal{A}(t)\right|$ for some $D\in\mathcal{D}_{n}$ and all $t\geq t_{0}$. Equivalently, it yields $D\mathcal{A}^{+}(t)D=\mathcal{A}^{+}(t)$ and $D\mathcal{A}^{-}(t)D=\left|\mathcal{A}^{-}(t)\right|$ for all $t\geq t_{0}$. It is easy to find $D\Delta_{\mathcal{A}^{+}(t)}D=\Delta_{\mathcal{A}^{+}(t)}$ and $D\Delta_{\left|\mathcal{A}^{-}(t)\right|}D=\Delta_{\left|\mathcal{A}^{-}(t)\right|}$ for all $t\geq t_{0}$. If we insert these facts into (\ref{Eq7}) and (\ref{Eq8}), then we can validate that $D\Phi_{\mathrm{even}}(t,t_{0})D=\Phi_{\mathrm{even}}(t,t_{0})$ and $D\Phi_{\mathrm{odd}}(t,t_{0})D=-\Phi_{\mathrm{odd}}(t,t_{0})$ for all $t\geq t_{0}$, respectively. Simultaneously, we consider $D\in\mathcal{D}_{n}$ and can define an orthogonal matrix $P=P^{-1}$ in the form of
\[
P=\begin{bmatrix}\frac{\Ds I+D}{\Ds2}&\frac{\Ds I-D}{\Ds2}\\\frac{\Ds I-D}{\Ds2}&\frac{\Ds I+D}{\Ds2}\end{bmatrix}.
\]

\noindent By noting the formulation of $\Psi(t,t_{0})$ in (\ref{Eq28}), we can obtain
\[
P\Psi(t,t_{0})P
=\begin{bmatrix}\Phi_{-\mathcal{L}_{\left|\mathcal{A}(t)\right|}}(t,t_{0})&0\\0&\Phi_{-\mathcal{L}_{\left|\mathcal{A}(t)\right|}}(t,t_{0})\end{bmatrix},~~~\forall t\geq t_{0}.
\]

\noindent This, together with $\lim_{t\to\infty}\Phi_{-\mathcal{L}_{\left|\mathcal{A}(t)\right|}}(t,t_{0})=1_{n}\nu^{\tp}(t_{0})$ exponentially fast, yields that both $\Phi_{\mathrm{odd}}(t,t_{0})$ and $\Phi_{\mathrm{even}}(t,t_{0})$ converge exponentially fast and
\begin{equation}\label{Eq14}
\aligned
\lim_{t\to\infty}\Phi_{\mathrm{odd}}(t,t_{0})
&=\frac{\Ds1_{n}\nu^{\tp}(t_{0})-D1_{n}\nu^{\tp}(t_{0})D}{\Ds2}\\
\lim_{t\to\infty}\Phi_{\mathrm{even}}(t,t_{0})
&=\frac{\Ds1_{n}\nu^{\tp}(t_{0})+D1_{n}\nu^{\tp}(t_{0})D}{\Ds2}
\endaligned,~~~\forall t_{0}\geq0.
\end{equation}

\noindent If we insert (\ref{Eq14}) into (\ref{Eq23}), then we can straightforwardly obtain
\[\lim_{t\to\infty}\Phi(t,t_{0})
=\lim_{t\to\infty}\Phi_{\mathrm{even}}(t,t_{0})-\lim_{t\to\infty}\Phi_{\mathrm{odd}}(t,t_{0})
=D1_{n}\nu^{\tp}(t_{0})D\]

\noindent exponentially fast.
\end{IEEEproof}

Unlike Theorem \ref{Thm3}, we can not directly use the convergence result of the stochastic matrix $\Phi_{-\mathcal{L}_{\left|\mathcal{A}(t)\right|}}(t,t_{0})$, $\forall t\geq t_{0}$ to solve the convergence problem of $\Phi(t,t_{0})$ associated with a s.s.ub. signed digraph $\mathcal{G}(t)$. To overcome this difficulty, we establish the following lemma on the relation between $\mathcal{G}(t)$ and $\mathbf{G}(t)$.

\begin{lem}\label{Lem9}
For any $t_{0}\geq0$, let Assumption \ref{Assum1} hold and the union $\bigcup_{p=1}^{M}\mathcal{G}_{\sigma_{p}}$ be strongly connected. If $\mathcal{G}(t)$ is s.s.ub., then $\mathbf{G}(t)$ is jointly strongly connected, namely, there exist some finite positive integer $h\in\mathbb{Z}$ and some corresponding sequence $\left\{k_{j}:j\in\mathbb{Z}_{+}\right\}$ with $k_{0}=0$ and $0<k_{j+1}-k_{j}\leq h$ such that the union $\bigcup_{p=k_{j}}^{k_{j+1}-1}\mathbf{G}_{p}$, $\forall j\in\mathbb{Z}_{+}$ of each collection of digraphs $\left\{\mathbf{G}_{k_{j}},\mathbf{G}_{k_{j}+1},\cdots,\mathbf{G}_{k_{j+1}-1}\right\}$ is strongly connected.
\end{lem}

\begin{IEEEproof}
By Assumption \ref{Assum1}, we consider the same sequence $\left\{k_{j}:j\in\mathbb{Z}_{+}\right\}$ in Lemma \ref{Lem1} and then denote $\mathcal{G}_{j}^{\cup}\triangleq\bigcup_{p=k_{j}}^{k_{j+1}-1}\mathcal{G}_{p}$, $\forall j\in\mathbb{Z}_{+}$. Thus, $\mathcal{G}_{j}^{\cup}=\bigcup_{p=1}^{M}\mathcal{G}_{\sigma_{p}}$, $\forall j\in\mathbb{Z}_{+}$ holds and is strongly connected. Because $\mathcal{G}(t)$ is s.s.ub., $\bigcup_{p=1}^{M}\mathcal{G}_{\sigma_{p}}=\bigcup_{t\geq t_{0}}\mathcal{G}(t)$ is structurally unbalanced. That is, each digraph $\mathcal{G}_{j}^{\cup}$, $\forall j\in\mathbb{Z}_{+}$ is structurally unbalanced. By following similar proof steps as in \cite[Lemma 2]{xcj:16}, we consider $\mathcal{G}_{j}^{\cup}$, $\forall j\in\mathbb{Z}_{+}$ and can deduce that each corresponding union $\mathbf{G}_{j}^{\cup}\triangleq\bigcup_{p=k_{j}}^{k_{j+1}-1}\mathbf{G}_{p}$, $\forall j\in\mathbb{Z}_{+}$ is strongly connected. Consequently, it is immediate to conclude the joint strong connectivity of $\mathbf{G}(t)$ in this lemma.
\end{IEEEproof}

Based on Lemmas \ref{Lem2} and \ref{Lem9} and with Theorems \ref{Thm4}-\ref{Thm1}, now we can present an exponential stability result for $\Phi(t,t_{0})$ under a s.s.ub. signed digraph $\mathcal{G}(t)$.

\begin{thm}\label{Thm6}
For any $t_{0}\geq0$, let Assumption \ref{Assum1} hold and the union $\bigcup_{p=1}^{M}\mathcal{G}_{\sigma_{p}}$ be strongly connected. If $\mathcal{G}(t)$ is s.s.ub., then $\lim_{t\to\infty}\Phi(t,t_{0})=0$ can be achieved and its limit is approached exponentially fast.
\end{thm}

\begin{IEEEproof}
By Lemma \ref{Lem9}, we consider $\Psi(t,t_{0})$ and the conventional digraph $\mathbf{G}(t)$, and then follow the same reason as used in the proof of $\lim_{t\to\infty}\Phi_{-\mathcal{L}_{\left|\mathcal{A}(t)\right|}}(t,t_{0})$ in Lemma \ref{Lem2} to arrive at (see also \cite{rb:05,mj:16})
\[\lim_{t\to\infty}\Psi(t,t_{0})=1_{2n}\mathbf{f}^{\tp}(t_{0})~\hbox{exponentially fast}
\]

\noindent for some $\mathbf{f}(t_{0})\geq0$ such that $\mathbf{f}^{\tp}(t_{0})1_{2n}=1$. Using the definition of $\Psi(t,t_{0})$ in (\ref{Eq28}), we can easily see that $\lim_{t\to\infty}\Phi_{\mathrm{odd}}(t,t_{0})$ and $\lim_{t\to\infty}\Phi_{\mathrm{even}}(t,t_{0})$ exist and satisfy
\begin{equation}\label{Eq15}
\lim_{t\to\infty}\Phi_{\mathrm{odd}}(t,t_{0})=\lim_{t\to\infty}\Phi_{\mathrm{even}}(t,t_{0})=\frac{\Ds1_{n}\nu^{\tp}(t_{0})}{\Ds2}~\hbox{exponentially fast}
\end{equation}

\noindent where we also incorporate the relation (\ref{Eq26}) of Theorem \ref{Thm2} and the result of Lemma \ref{Lem2}. Based on the relation (\ref{Eq23}) of Theorem \ref{Thm4}, the exponential stability in this theorem is a straightforward consequence of (\ref{Eq15}).
\end{IEEEproof}

\begin{rem}\label{Rem6}
In Theorem \ref{Thm6}, $\mathbf{f}(t_{0})=\left[\nu^{\tp}(t_{0}),\nu^{\tp}(t_{0})\right]^{\tp}/2$ clearly holds with the vector $\nu(t_{0})$ in Lemma \ref{Lem2}. This, together with Theorem \ref{Thm3}, demonstrates that there exists an inherent relationship between the convergence of $\Psi(t,t_{0})$ and of $\Phi_{-\mathcal{L}_{\left|\mathcal{A}(t)\right|}}(t,t_{0})$ for any signed digraph $\mathcal{G}(t)$. Actually, it is consistent with the fact that $\Psi(t,t_{0})$ converges if and only if both $\Phi(t,t_{0})$ and $\Phi_{-\mathcal{L}_{\left|\mathcal{A}(t)\right|}}(t,t_{0})$ converge. Such a relationship is bridged with the introductions of $\Phi_{\mathrm{odd}}(t,t_{0})$ and $\Phi_{\mathrm{even}}(t,t_{0})$. But, note that though both $\Phi_{\mathrm{odd}}(t,t_{0})$ and $\Phi_{\mathrm{even}}(t,t_{0})$ converge exponentially fast, we have different convergence results (\ref{Eq14}) and (\ref{Eq15}) when $\mathcal{G}(t)$ is s.s.b. and s.s.ub., respectively.
\end{rem}

Since the simultaneous structural balance of $\mathcal{G}(t)$ is mutually exclusive with the simultaneous structural unbalance of $\mathcal{G}(t)$, we can strengthen the results of Theorems \ref{Thm3} and \ref{Thm6} to present a theorem that connects the convergence of $\Phi(t,t_{0})$ equivalently with the simultaneous structural property of $\mathcal{G}(t)$.

\begin{thm}\label{Thm7}
For any $t_{0}\geq0$, let Assumption \ref{Assum1} hold and the union $\bigcup_{p=1}^{M}\mathcal{G}_{\sigma_{p}}$ be strongly connected. Then $\Phi(t,t_{0})$ converges exponentially fast to some matrix $\Phi_{\infty}(t_{0})$, where $\Phi_{\infty}(t_{0})=\left(D1_{n}\right)\left[\nu^{\tp}(t_{0})D\right]$ (respectively, $\Phi_{\infty}(t_{0})=0$) if and only if $\mathcal{G}(t)$ is s.s.b. (respectively, s.s.ub.).
\end{thm}

\begin{IEEEproof}
A consequence of Theorems \ref{Thm3} and \ref{Thm6} as well as the mutually exclusive properties between simultaneous structural balance and unbalance.
\end{IEEEproof}

Analogous to Theorem \ref{Thm7}, the following theorem presents an exponential convergence result for opinion dynamics in signed networks, which is tied closely to the simultaneous structural balance property of switching signed digraphs.

\begin{thm}\label{Thm5}
For any $t_{0}\geq0$, let Assumption \ref{Assum1} hold and the union $\bigcup_{p=1}^{M}\mathcal{G}_{\sigma_{p}}$ be strongly connected. Then the system (\ref{Eq3}) achieves bipartite consensus (respectively, stability) if and only if $\mathcal{G}(t)$ is s.s.b. (respectively, s.s.ub.), where the convergence possesses an exponentially fast speed.
\end{thm}

\begin{IEEEproof}
A straightforward consequence of Theorem \ref{Thm7}.
\end{IEEEproof}

\begin{rem}\label{Rem2}
In Theorem \ref{Thm5}, we present a bipartite consensus result of signed networks, together with a necessary and sufficient guarantee related to the s.s.b. property of switching signed digraphs. From Lemma \ref{Lem1}, this convergence result works for signed networks under switching signed digraphs that are jointly strongly connected. Although similar results have been developed in, e.g., \cite{xcj:16}, Theorem \ref{Thm5} is proposed to disclose a close relationship between bipartite consensus of opinion dynamics and simultaneous structural balance of switching signed digraphs.
\end{rem}

\section{Conclusions and Remarks}\label{sec7}

In this paper, the dynamics evolution of opinions in signed networks has been discussed. We have considered the class of Laplacian flows and have established a state transition matrix-based approach. It achieves the separation of cooperations and antagonisms, which also helps to connect opinion dynamics in signed networks to those in conventional networks. Moreover, the proposed approach is applicable to opinion dynamics under general time-varying signed digraphs that need no assumptions on connectivity, structural balance or digon sign-symmetry. It is particularly applied to exploit convergence results of opinion dynamics in signed networks associated with switching signed digraphs, where bipartite consensus (or stability) is achieved if and only if the simultaneous structural balance (or unbalance) can be ensured.

Since the proposed state transition matrix-based approach is general and can bridge a clear relationship between the opinion dynamics problem with antagonisms and an associated opinion dynamics problem under a conventional digraph with a specific structure and twice enlarged nodes, it is applicable to obtaining many more results of opinion dynamics associated with signed digraphs, especially not limited to the case of switching signed digraphs considered in this paper. For example, it may be used in convergence analysis of opinion dynamics in the presence of uniformly strongly connected, cut-balanced or type-symmetric time-varying signed digraphs, like \cite{h:14,pmc:16}. Moreover, there are many aspects that have not been discussed in this paper, including opinions possessing the general linear/nonlinear dynamics, finite/fixed-time convergence rate, or influences caused by communication delays or noises. Also, the discrete-time counterpart problems have not been considered. These will be our future topics.

%
\appendices

\section{Proof of lemma \ref{Lem8}}\label{ap1}

\begin{IEEEproof}
We first prove that $\Phi_{-\mathcal{L}_{\mathcal{A}^{+}(t)}-\Delta_{\left|\mathcal{A}^{-}(t)\right|}}(t,t_{0})$, $\forall t\geq t_{0}$ is a nonnegative matrix. Towards this end, we denote $a(t)=\max_{i\in\mathcal{I}_{n}}\sum_{j=1}^{n}\left|a_{ij}(t)\right|$. The use of (\ref{Eq1}) leads to
\[\aligned
\Xi(t)
&=a(t)I-\mathcal{L}_{\mathcal{A}^{+}(t)}-\Delta_{\left|\mathcal{A}^{-}(t)\right|}\\
&=\left[a(t)I-\Delta_{\left|\mathcal{A}(t)\right|}\right]+\mathcal{A}^{+}(t)\\
&\geq0,~~\forall t\geq t_{0}
\endaligned\]

\noindent and thus $\Phi_{\Xi(t)}(t,t_{0})\geq I$, $\forall t\geq t_{0}$. Similarly, for the scalar case, we have $\Phi_{a(t)}(t,t_{0})\geq1$, $\forall t\geq t_{0}$. This ensures $\Phi_{-a(t)}(t,t_{0})=\Phi_{a(t)}^{-1}(t,t_{0})>0$, $\forall t\geq t_{0}$. By inserting these facts and using $-\mathcal{L}_{\mathcal{A}^{+}(t)}-\Delta_{\left|\mathcal{A}^{-}(t)\right|}=-a(t)I+\Xi(t)$, we can deduce
\[\aligned
\Phi_{-\mathcal{L}_{\mathcal{A}^{+}(t)}-\Delta_{\left|\mathcal{A}^{-}(t)\right|}}(t,t_{0})
&=\Phi_{-a(t)I}(t,t_{0})\Phi_{\Xi(t)}(t,t_{0})\\
&=\Phi_{-a(t)}(t,t_{0})\Phi_{\Xi(t)}(t,t_{0})\\
&\geq0,~~\forall t\geq t_{0}.
\endaligned\]

\noindent Then with $\mathcal{L}_{\mathcal{A}^{+}(t)}1_{n}=0$ and $\Delta_{\left|\mathcal{A}^{-}(t)\right|}\geq0$, we can derive
\[\aligned
\Xi(t)1_{n}
&=a(t)1_{n}-\mathcal{L}_{\mathcal{A}^{+}(t)}1_{n}-\Delta_{\left|\mathcal{A}^{-}(t)\right|}1_{n}\\
&=a(t)1_{n}-\Delta_{\left|\mathcal{A}^{-}(t)\right|}1_{n}\\
&\leq a(t)1_{n},~~\forall t\geq t_{0}
\endaligned\]

\noindent which, together with $a(t)>0$ and $\Xi(t)\geq0$, $\forall t\geq t_{0}$, results in
\[\aligned
\left[\Xi(\theta_{1})\Xi(\theta_{2})\cdots\Xi(\theta_{k})\right]1_{n}
&\leq\left[a(\theta_{1})a(\theta_{2})\cdots a(\theta_{k})\right]1_{n},~~\forall\theta_{i}\geq0,i=1,2,\cdots,k.
\endaligned\]

\noindent By considering the Peano-Baker series expressions of $\Phi_{\Xi(t)}(t,t_{0})$ and $\Phi_{a(t)}(t,t_{0})$, we can obtain
\[
\aligned
\Phi_{\Xi(t)}(t,t_{0})1_{n}
&=1_{n}
+\sum_{k=1}^{\infty}\left[\int_{t_{0}}^{t}\Xi(\theta_{1})\int_{t_{0}}^{\theta_{1}}\Xi(\theta_{2})\cdots\right.\\
&~~~~~~~~~~~~~\left.\times\int_{t_{0}}^{\theta_{k-1}}\Xi(\theta_{k})d\theta_{k}\cdots d\theta_{2}d\theta_{1}\right]1_{n}\\
&\leq1_{n}
+\sum_{k=1}^{\infty}\left[\int_{t_{0}}^{t}a(\theta_{1})\int_{t_{0}}^{\theta_{1}}a(\theta_{2})\cdots\right.\\
&~~~~~~~~~~~~~\left.\times\int_{t_{0}}^{\theta_{k-1}}a(\theta_{k})d\theta_{k}\cdots d\theta_{2}d\theta_{1}\right]1_{n}\\
&=\Phi_{a(t)}(t,t_{0})1_{n},~~\forall t\geq t_{0}.
\endaligned
\]

\noindent Consequently, we can employ \cite[eq. (8.1.11)]{hj:85} to arrive at
\[\aligned
\Phi_{-\mathcal{L}_{\mathcal{A}^{+}(t)}-\Delta_{\left|\mathcal{A}^{-}(t)\right|}}(t,t_{0})1_{n}
&=\Phi_{-a(t)}(t,t_{0})\left[\Phi_{\Xi(t)}(t,t_{0})1_{n}\right]\\
&\leq\Phi_{-a(t)}(t,t_{0})\left[\Phi_{a(t)}(t,t_{0})1_{n}\right]\\
&=1_{n},~~\forall t\geq t_{0}.
\endaligned\]

\noindent Thus, $\Phi_{-\mathcal{L}_{\mathcal{A}^{+}(t)}-\Delta_{\left|\mathcal{A}^{-}(t)\right|}}(t,t_{0})\geq0$ is a substochastic matrix.

To prove (\ref{Eq6}), let us denote $\Theta(t,t_{0})$ as the matrix function on the right hand side of it. By incorporating (\ref{Eq4}), we can validate
\[\aligned
\frac{\Ds d\left[\Theta(t,t_{0})-\Phi(t,t_{0})\right]}{\Ds dt}
&=\left[-\mathcal{L}_{\mathcal{A}^{+}(t)}-\Delta_{\left|\mathcal{A}^{-}(t)\right|}\right]\left[\Theta(t,t_{0})-\Phi(t,t_{0})\right],~~\forall t\geq t_{0}.
\endaligned\]

\noindent With (\ref{Eq6}), we can see $\Theta(t_{0},t_{0})=\Phi_{-\mathcal{L}_{\mathcal{A}^{+}(t)}-\Delta_{\left|\mathcal{A}^{-}(t)\right|}}(t_{0},t_{0})=I$, which yields $\Theta(t_{0},t_{0})-\Phi(t_{0},t_{0})=0$. By solving the above linear differential matrix equation, we can derive
\[\aligned
\Theta(t,t_{0})-\Phi(t,t_{0})
&=\Phi_{-\mathcal{L}_{\mathcal{A}^{+}(t)}-\Delta_{\left|\mathcal{A}^{-}(t)\right|}}(t,t_{0})\left[\Theta(t_{0},t_{0})-\Phi(t_{0},t_{0})\right]\\
&=0,~~\forall t\geq t_{0}
\endaligned\]

\noindent i.e., (\ref{Eq6}) holds.
\end{IEEEproof}


\end{document}